\documentclass[fp]{jpsj2}
\usepackage{txfonts}
\usepackage{bm}
\title{ 
Momentum-Dependent Local Ansatz Approach to the Quasiparticle States
}

\author{
Yoshiro \textsc{Kakehashi}\thanks{yok@sci.u-ryukyu.ac.jp, to be published in J. Phys. Soc. Jpn.}
}

\inst{
Department of Physics and Earth Sciences,
Faculty of Science, \\
University of the Ryukyus, \\
1 Senbaru, Nishihara, Okinawa, 903-0213, Japan
}

\abst{\vspace*{10mm}
Momentum-dependent local ansatz (MLA) wavefunction approach to correlated electrons in  solids has been extended to the quasiparticle excited states on the basis of the Fermi liquid picture.  The quasiparticle energy is derived from the MLA-base excited state wavefunction within the single-site approximation.  The quasiparticle weight and quasiparticle energy have been calculated for the Hubbard model on the hypercubic lattice in infinite dimensions as well as the simple cubic lattice. By comparing the numerical results with the single-particle excitation spectra obtained by the Green function method, it is shown that the MLA excited  wavefunctions describe well the quasiparticle states in the weak and intermediate Coulomb interaction regime in the high dimensional system.
}

\kword{quasiparticle state, quasiparticle energy, wavefunction method, electron correlations, Hubbard model}

\begin{document}
\maketitle

\section{Introduction} 
The wavefunction method has been a simple and useful approach to understanding the ground-state property of correlated electrons in solids such as the magnetism, the metal-insulator transition, the heavyfermion behavior, and the super 
conductivity \cite{fulde95,fulde12,fazekas99,gebhard97,kake12}.

Among various wavefunctions, the Gutzwiller ansatz wavefunction (GA) is one of the popular wavefunctions for correlated electrons \cite{gutz63,gutz64,gutz65,br70,bune98,bune03-1,lana17,lana21}.  The GA describes electron correlations by suppressing the amplitudes of doubly-occupied electron states in the Hartree-Fock wavefunction.  Using the GA and the single-band Hubbard model, Gutzwiller \cite{gutz63,gutz64,gutz65} clarified the role of electron correlations in the metallic ferromagnetism.
Stollhoff and Fulde \cite{stoll77,stoll78,stoll80} proposed the local ansatz wavefunction (LA) taking into account the Hilbert space expanded by the residual Coulomb interactions, and discussed the electron correlations in transition metals.  The LA is an approach from the weakly correlated limit and allowed us to apply a realistic tight-binding multi-orbital model.  Baeriswyl \cite{baer87,baer00} proposed a wavefunction which takes into account the Hilbert space expanded by the hopping operators, starting from the atomic wavefunction in order to describe electron correlations in the strong Coulomb interaction regime.
There are many wavefunctions which explicitly take into account nonlocal electron correlations \cite{hetenyi10,jast55,kaplan82,yokoyama90,tahara08}. These wavefunctions have been applied to the strongly correlated electron systems with use of the numerical techniques such as the variational Monte-Carlo method \cite{ceperley77,yokoyama87}.

Although the variational wavefunctions mentioned above have clarified many aspects of electron correlations, they do not reduce to the second-order perturbation theory in the weak Coulomb interaction limit.  Thus they are not suitable for the quantitative description of correlated electrons in the weak Coulomb interaction regime.  We therefore proposed the momentum-dependent local ansatz wavefunction (MLA) which works best in the weak and intermediate Coulomb interaction regime \cite{kake08,pat11,pat13,pat13-2}.

The MLA is an extension of the LA.   It reproduces the second order perturbation theory in the weak Coulomb interaction limit and interpolates between the weak and strong interaction limits by introducing momentum dependent variational parameters.  The theory explained a large mass enhancement of $\beta$-Mn \cite{kake22}.  Moreover, we extended the MLA to the first-principles version \cite{kake16,chan16,kake16-2,kake21} using the tight-binding LDA+U Hamiltonian which is based on the density functional band theory with the local density approximation (LDA).  The first-principles MLA clarified the quantitative aspects of the ground-state properties of the iron-group transition metals.

In the present paper, we extend the MLA approach to the quasiparticle (QP) excited states.  The QP state with momentum $k$ and spin $\sigma$ in the vicinity of the Fermi level is adiabatically connected to its noninteracting counterpart with the same $k$ and the same spin according to the Fermi liquid picture \cite{land56}.  We will demonstrate that the quasiparticle excited states can be constructed from the Hartree-Fock excited states using the same MLA correlator as used at the ground state.  
We call the MLA wavefunction for the QP excited states the QPMLA hereafter.
The same type of excitation theories using the LA and GA have been proposed for various systems \cite{hors79,hors83,bune03-2}.  These theories, however, do not yield the correct result in the weak Coulomb interaction limit.
The QP energy in the MLA approach yields the second order perturbation theory in the weak Coulomb interaction limit. 
We also show that the QPMLA describes well the excitations in the intermediate Coulomb interaction regime in the high dimensional system, comparing the numerical results of the QP energy with those of the excitation spectra obtained by the Green function method \cite{kake04-1}.  

In the following section, Sect. 2.1, we adopt the single-band Hubbard model and briefly review the ground-state MLA.  In Sect. 2.2, we extend the MLA to the QP excited states according to the Fermi liquid picture, and derive the QP energy in the single-site approximation (SSA).  In Sect. 3, we present the numerical results of the QP weight for the Hubbard model on the hypercubic lattice in infinite dimensions as well as the QP energy curves along the high symmetry lines in the first Brillouin zone of the simple cubic lattice.  By comparing the results with those obtained from the Green function method, we will show that the theory describes well the QP states in the weak and intermediate Coulomb interaction regime.  In the last Sect. 4, we summarize the present work and discuss the remaining problems towards future investigations.

\section{Momentum-Dependent Local Ansatz Approach}
\subsection{Ground-state MLA}

We briefly review in this subsection the momentum-dependent local ansatz (MLA) approach for the ground state \cite{kake08,pat11,pat13,pat13-2}.  
We consider the single-band Hubbard model with an atom per unit cell as follows. 
\begin{eqnarray}
H = \sum_{i \sigma} \tilde{\epsilon}_{0} n_{i\sigma} 
+ \sum_{ij \sigma} t_{i j} \, a_{i \sigma}^{\dagger} a_{j \sigma} 
+ U \sum_{i} \, n_{i \uparrow} n_{i \downarrow} \ .
\label{hub}
\end{eqnarray}
Here $\tilde{\epsilon}_{0}$ is the atomic level measured from the Fermi level $\epsilon_{\rm F}$; $\tilde{\epsilon}_{0} = \epsilon_{0} - \epsilon_{\rm F}$, $\epsilon_{0}$ being the original atomic level.
$t_{ij}$ is the transfer integral between sites $i$ and $j$.  $U$
is the intra-atomic Coulomb interaction energy parameter.  $ a_{i \sigma}^{\dagger}$
($ a_{i \sigma}$) denotes the creation (annihilation) operator for an
electron on site $i$ with spin $\sigma$, and 
$n_{i\sigma}=a_{i\sigma}^{\dagger} a_{i \sigma}$ is the electron density
operator on site $i$ for spin $\sigma$.

In the Hartree-Fock approximation, we replace the Hamiltonian (\ref{hub}) 
with an independent-electron Hamiltonian as follows.
\begin{eqnarray}
H_{0} = \sum_{i \sigma} \epsilon_{\sigma} n_{i\sigma} + 
\sum_{ij \sigma} t_{i j} \, a_{i \sigma}^{\dagger} a_{j \sigma} 
- U \sum_{i} \, \langle n_{i \uparrow} \rangle_{0} 
\langle n_{i\downarrow} \rangle_{0} \ .
\label{hf}
\end{eqnarray}
Here $\epsilon_{\sigma}=\epsilon_{0} + U \langle n_{i -\sigma} \rangle_{0} - \epsilon_{\rm F}$, and $\langle \sim \rangle_{0}$ denotes the average with respect to the Hartree-Fock ground state.   Note that in the Hartree-Fock approximation the effects of the Coulomb interactions appear only via the Hartree-Fock potential $U \langle n_{i -\sigma} \rangle_{0}$ and the double counting term at the end of the right-hand-side (rhs) of Eq. (\ref{hf}). 

The Hartree-Fock Hamiltonian  (\ref{hf}) in the paramagnetic or ferromagnetic state is diagonalized in the momentum representation as follows.
\begin{eqnarray}
H_{0} = \sum_{k \sigma} \epsilon_{k\sigma} n_{k\sigma}  
- U \sum_{i} \, \langle n_{i \uparrow} \rangle_{0} 
\langle n_{i\downarrow} \rangle_{0} \ .
\label{hfdia}
\end{eqnarray}
Here $\epsilon_{k\sigma}$ is the momentum representation of the Hartree-Fock one-electron Hamiltonian $H_{ij} = \epsilon_{\sigma} \delta_{ij} + t_{ij}$, and $n_{k\sigma} = a_{k\sigma}^{\dagger} a_{k \sigma}$ is the electron density operator for an electron with momentum $k$ and spin $\sigma$.  $a_{k\sigma}^{\dagger}$ ($a_{k \sigma}$) is the creation (annihilation) operator for an electron with momentum $k$ and spin $\sigma$.

The original Hamiltonian (\ref{hub}) is expressed with use of the
Hartree-Fock Hamiltonian (\ref{hf}) and the residual Coulomb interactions $H_{\rm I}$ as follows.
\begin{eqnarray}
H = H_{0} + H_{\rm I} = H_{0} + \sum_{i} \, U O_{i} \ .
\label{hub2}
\end{eqnarray}
Here $H_{\rm I} = \sum_{i} \, U O_{i}$, 
$O_{i}=\delta n_{i \uparrow}\delta n_{i \downarrow} $, and 
$\delta n_{i\sigma} = n_{i\sigma} - \langle n_{i\sigma} \rangle_{0}$.

In the MLA the ground-state wavefunction $| \Psi_{0} \rangle$ is assumed to be given by
\begin{eqnarray}
|\Psi_{0} \rangle = \big[ \prod_{i} (1 - \tilde{O}_{i}) \big] |\phi \rangle \ .
\label{mla}
\end{eqnarray}
Here $|\phi \rangle$ denotes the Hartree-Fock ground state. 
Operator $\tilde{O}_{i}$ is a local correlator on site $i$ with momentum-dependent variational parameters $\eta_{k^{\prime}_{2}k_{2}k^{\prime}_{1}k_{1}}$.
\begin{eqnarray}
\tilde{O}_{i} = \sum_{k_{1}k_{2}k^{\prime}_{1}k^{\prime}_{2}} 
\langle k^{\prime}_{1}|i \rangle \langle i|k_{1} \rangle 
\langle k^{\prime}_{2}|i \rangle \langle i|k_{2} \rangle
\eta_{k^{\prime}_{2}k_{2}k^{\prime}_{1}k_{1}} 
\delta(a^{\dagger}_{k^{\prime}_{2}\downarrow}a_{k_{2}\downarrow})
\delta(a^{\dagger}_{k^{\prime}_{1}\uparrow}a_{k_{1}\uparrow}) \ .
\label{otilde}
\end{eqnarray}
Here $\langle i|k \rangle$ is the overlap matrix between the local orbital $i$ and the state for an electron with momentum $k$, $a_{k\sigma} = \sum_{i} \langle k|i \rangle a_{i\sigma}$, and 
$\delta(a^{\dagger}_{k^{\prime}\sigma}a_{k\sigma}) = a^{\dagger}_{k^{\prime}\sigma}a_{k\sigma} - \langle a^{\dagger}_{k^{\prime}\sigma}a_{k\sigma} \rangle_{0}$.
Note that operator $\tilde{O}_{i}$ reduces to a local operator $O_{i} = \delta n_{i\uparrow} \delta n_{i\downarrow}$ when the variational parameters are momentum independent: $\eta_{k^{\prime}_{2}k_{2}k^{\prime}_{1}k_{1}} = 1$, and the wavefunction (\ref{mla}) reduces to the LA wavefunction \cite{stoll77,stoll78,stoll80}.
We also note that Eq. (\ref{mla}) implies that the operator
\begin{eqnarray}
\Omega_{\rm MLA} = \prod_{i} (1 - \tilde{O}_{i}) \ ,
\label{mlaop}
\end{eqnarray}
is an approximate waveoperator which transforms the Hartree-Fock ground state $|\phi \rangle$ into the correlated ground state $|\Psi_{0} \rangle$.

The variational parameters are determined by minimizing the ground-state energy $E$.
\begin{eqnarray}
E = \epsilon_{\rm F} N + \langle H \rangle_{0} + L \epsilon_{\rm c} \ . 
\label{eg}
\end{eqnarray}
Here $N$ ($L$) is the total number of electrons (lattice sites).  $\langle H \rangle_{0}$ denotes the Hartree-Fock ground-state energy given by
\begin{eqnarray}
\langle H \rangle_{0} = \langle H_{0} \rangle_{0} = \sum^{\rm occ}_{k \sigma} \epsilon_{k\sigma}  
- U \sum_{i} \, \langle n_{i \uparrow} \rangle_{0} 
\langle n_{i\downarrow} \rangle_{0} \ .
\label{eghf}
\end{eqnarray}
The last term of Eq. (\ref{eg}) is the correlation correction and $\epsilon_{\rm c}$ is the correlation energy per site.
\begin{eqnarray}
\epsilon_{\rm c} 
= \frac{1}{L} \dfrac{\langle \Psi_{0} |\tilde{H}| \Psi_{0} \rangle}{\langle \Psi_{0} | \Psi_{0} \rangle} 
= \frac{1}{L} \dfrac{\langle \Omega^{\dagger}_{\rm MLA} \tilde{H} \Omega_{\rm MLA} \rangle_{0}}{\langle \Omega^{\dagger}_{\rm MLA} | \Omega_{\rm MLA} \rangle_{0}} \ .
\label{corr}
\end{eqnarray}
Here $\tilde{H}=H - \langle H \rangle_{0}$.

In the calculations of the correlation energy, we make use of the single-site approximation (SSA), which is given for a physical quantity $\tilde{A}= A-\langle A \rangle_0$ as follows:
\begin{eqnarray}
\dfrac{\langle \Omega^{\dagger}_{\rm MLA} \tilde{A} \Omega_{\rm MLA} \rangle_{0}}{\langle \Omega^{\dagger}_{\rm MLA} | \Omega_{\rm MLA} \rangle_{0}} \approx 
\sum^{L}_{i} \dfrac{\langle R^{\dagger}_{i} \tilde{A} R_{i} \rangle_{0}}
{\langle R^{\dagger}_{i} R_{i} \rangle_{0}} \ .
\label{ssa}
\end{eqnarray}
Here $R_{i} = 1 - \tilde{O}_{i}$.  The above formula of the SSA has been given in Appendix A of our previous paper \cite{kake08}.
Adopting the SSA, we obtain the correlation energy as follows.
\begin{eqnarray}
\epsilon_{\rm c} = \dfrac{-\langle
 \tilde{O}^{\dagger}_{i}\tilde{H}\rangle_{0} -
\langle \tilde{H} \tilde{O}_{i} \rangle_{0} + 
\langle \tilde{O}^{\dagger}_{i}\tilde{H}\tilde{O}_{i}\rangle_{0}}
{1 + \langle \tilde{O}^{\dagger}_{i}\tilde{O}_{i} \rangle_{0}} \ .
\label{ecssa}
\end{eqnarray}
Each term at the rhs of Eq. (\ref{ecssa}) can be calculated with use of the Wick theorem \cite{kake08}.
Moreover, in order to simplify the calculations of the correlation energy, we make use of the interpolated form of the variational parameters, which is exact in the weak and atomic Coulomb interaction limits,
\begin{eqnarray}
\eta_{k^{\prime}_{2}k_{2}k^{\prime}_{1}k_{1}} = 
\dfrac{U\tilde{\eta}}
{\epsilon_{k^{\prime}_{2\downarrow}} - \epsilon_{k_{2\downarrow}} + \epsilon_{k^{\prime}_{1\uparrow}} - \epsilon_{k_{1\uparrow}} - \epsilon_{\rm c}} \ .
\label{etaint}
\end{eqnarray}
Here $\tilde{\eta}$ is a momentum independent variational parameter.

Substituting Eq. (\ref{etaint}) into Eq. (\ref{ecssa}), we obtain the explicit expressions for each term at the rhs of Eq. (\ref{ecssa}) as follows.
\begin{eqnarray}
\langle \tilde{H} \tilde{O}_{i} \rangle_{0} = 
\langle \tilde{O}^{\dagger}_{i}\tilde{H} \rangle^{*}_{0}
=A U^{2} \tilde{\eta} \ , \label{ecterm1} \\
\langle \tilde{O}^{\dagger}_{i}\tilde{H}\tilde{O}_{i}\rangle_{0} = B U^{2} \tilde{\eta}^{2} 
\ , \hspace{12mm}  \label{ecterm2} \\
\langle \tilde{O}^{\dagger}_{i}\tilde{O}_{i} \rangle_{0} = C U^{2}  \tilde{\eta}^{2} 
\ . \hspace{16mm}
\label{ecterm3}
\end{eqnarray}
The coefficients $A$, $B=B^{(1)}+UB^{(2)}$, and $C$ are given as follows.
\begin{align} 
A =
\int \frac{\Big[ \prod \limits^4_{n=1} d\epsilon_{n}\Big] 
X_{\uparrow}(\epsilon_{1}, \epsilon_{2}, \epsilon_{3}, \epsilon_{4})}
{\epsilon_{4} - \epsilon_{3} + \epsilon_{2} - \epsilon_{1} -\epsilon_{\rm c}} \ ,
\label{hoa}
\end{align}
\begin{align}
B^{(1)} = 
\int \frac{\Big[ \prod \limits^4_{n=1} d\epsilon_{n}\Big]
X_{\uparrow}(\epsilon_{1}, \epsilon_{2}, \epsilon_{3}, \epsilon_{4}) 
(\epsilon_{4} - \epsilon_{3} + \epsilon_{2} - \epsilon_{1})} 
{(\epsilon_{4} - \epsilon_{3} + \epsilon_{2} - \epsilon_{1} -\epsilon_c)^{2}} \ ,
\label{ohob1}
\end{align}
\begin{align}
B^{(2)} &=  
\int \frac{\Big[ \prod \limits^4_{n=1} d\epsilon_{n}\Big] 
X_{\uparrow}(\epsilon_{1}, \epsilon_{2}, \epsilon_{3}, \epsilon_{4})}
{\epsilon_{4} - \epsilon_{3} + \epsilon_{2} - \epsilon_{1} -\epsilon_{\rm c}} 
\int d\epsilon_{5} d\epsilon_{6} \rho_{\uparrow}(\epsilon_{5}) \rho_{\downarrow}(\epsilon_{6}) \nonumber \\ 
& \hspace*{5mm} \times 
\Bigg( \frac{f(\epsilon_{5}) f(\epsilon_{6})}
{\epsilon_{4} - \epsilon_{6} + \epsilon_{2} - \epsilon_{5} -\epsilon_{\rm c}} 
- \frac{ f(-\epsilon_{5}) f(\epsilon_{6})}
{\epsilon_{4} - \epsilon_{6} + \epsilon_{5} - \epsilon_{1} -\epsilon_{\rm c}}  \nonumber \\
& \hspace*{12mm}  - \frac{f(\epsilon_{5}) f(-\epsilon_{6})}
{\epsilon_{6} - \epsilon_{3} + \epsilon_{2} - \epsilon_{5} -\epsilon_{\rm c}} 
+ \frac{ f(-\epsilon_{5}) f(-\epsilon_{6})}
{\epsilon_{6} - \epsilon_{3} + \epsilon_{5} - \epsilon_{1} -\epsilon_{\rm c}} \Bigg) \ ,
\label{ohob2}
\end{align}
\begin{align} 
C =  
\int \frac{ \Big[ \prod \limits^4_{n=1} d\epsilon_{n} \Big]
X_{\uparrow}(\epsilon_{1}, \epsilon_{2}, \epsilon_{3}, \epsilon_{4})} 
{(\epsilon_{4} - \epsilon_{3} + \epsilon_{2} - \epsilon_{1} -\epsilon_c)^{2}} \  ,
\label{ooc}
\end{align}
and 
\begin{eqnarray}
X_{\sigma}(\epsilon_{1}, \epsilon_{2}, \epsilon_{3}, \epsilon_{4}) = 
\rho_{\sigma}(\epsilon_{1}) \rho_{\sigma}(\epsilon_{2}) \rho_{-\sigma}(\epsilon_{3}) \rho_{-\sigma}(\epsilon_{4}) f(\epsilon_{1}) f(-\epsilon_{2}) f(\epsilon_{3}) f(-\epsilon_{4}) \ .
\label{xsigma}
\end{eqnarray}
Here $\rho_{\sigma}(\epsilon)$ is the local density of states defined by 
$\rho_{\sigma}(\epsilon) = \sum_{k} |\langle i | k \rangle|^{2} \delta(\epsilon - \epsilon_{k\sigma})$, and $f(\epsilon)$ is the Fermi distribution function at the zero temperature.

Substituting Eqs. (\ref{ecterm1}), (\ref{ecterm2}), and (\ref{ecterm3}) into Eq. (\ref{ecssa}), we obtain the correlation energy as follows.
\begin{eqnarray}
\epsilon_{\rm c} = \dfrac{ -2A U^{2} \tilde{\eta}  +  B U^{2} \tilde{\eta}^{2}}
{1 + C U^{2}  \tilde{\eta}^{2}} \ .
\label{ecssa2}
\end{eqnarray}
Note that the ground state energy (\ref{eg}) satisfies the variational principle:
\begin{eqnarray}
E \ge E_{0} \ ,
\label{vp}
\end{eqnarray}
where $E_{0}$ is the exact ground state energy.  The variational parameter $\tilde{\eta}$ in Eq. (\ref{ecssa2}) is therefore obtained from the stationary condition $\delta E = L \delta \epsilon_{\rm c} = 0$ as
\begin{eqnarray}
\tilde{\eta} = \frac{-B + \sqrt{B^{2} + 4ACU^{2}}}{2ACU^{2}} \ .
\label{etat2}
\end{eqnarray}

The electron number per site $n$ is also obtained in the same way as
\begin{eqnarray}
n = \langle n_{i} \rangle = \langle n_{i} \rangle_{0} + 
\dfrac{D U^{2} \tilde{\eta}^{2}}
{1 + C U^{2}  \tilde{\eta}^{2}} \ .
\label{ni}
\end{eqnarray}
Here $\langle n_{i} \rangle_{0}$ is the Hartree-Fock electron number per site, and
\begin{align}
D &=  
\int \frac{\Big[ \prod \limits^4_{n=1} d\epsilon_{n}\Big] 
X_{\uparrow}(\epsilon_{1}, \epsilon_{2}, \epsilon_{3}, \epsilon_{4})}
{\epsilon_{4} - \epsilon_{3} + \epsilon_{2} - \epsilon_{1} -\epsilon_{\rm c}}  \nonumber \\ 
& \hspace*{5mm} \times
\int d\epsilon_{5}  
\Bigg( \frac{\rho_{\uparrow}(\epsilon_{5}) f(-\epsilon_{5})}
{\epsilon_{4} - \epsilon_{3} + \epsilon_{5} - \epsilon_{1} -\epsilon_{\rm c}} 
- \frac{\rho_{\uparrow}(\epsilon_{5}) f(\epsilon_{5})}
{\epsilon_{4} - \epsilon_{3} + \epsilon_{2} - \epsilon_{5} -\epsilon_{\rm c}}  \nonumber \\
& \hspace*{20mm}  + \frac{\rho_{\downarrow}(\epsilon_{5})f(-\epsilon_{5})}
{\epsilon_{5} - \epsilon_{3} + \epsilon_{2} - \epsilon_{1} -\epsilon_{\rm c}} 
- \frac{\rho_{\downarrow}(\epsilon_{5}) f(\epsilon_{5})}
{\epsilon_{4} - \epsilon_{5} + \epsilon_{2} - \epsilon_{1} -\epsilon_{\rm c}} \Bigg) \ .
\label{ncd}
\end{align}
Equations (\ref{ecssa2}), (\ref{etat2}), and (\ref{ni}) form the self-consistent equations for $\tilde{\eta}$, $\epsilon_{\rm c}$, and $\epsilon_{\rm F}$.
Note that the electronic structure of the system is taken into account via the local density of states $\rho_{\sigma}(\epsilon)$ in the present theory.

Momentum distribution function $\langle n_{k\sigma} \rangle$ is given in the SSA as follows.
\begin{eqnarray}
\langle n_{k\sigma} \rangle = f(\epsilon_{k\sigma}) + 
\dfrac{\lambda_{1-\sigma}(\epsilon_{k\sigma})f(-\epsilon_{k\sigma}) -
\lambda_{2-\sigma}(\epsilon_{k\sigma})f(\epsilon_{k\sigma})}
{1 + C U^{2}  \tilde{\eta}^{2}} \ .
\label{nk}
\end{eqnarray}
The functions $\lambda_{1-\sigma}(\epsilon_{k\sigma})$ and 
$\lambda_{2-\sigma}(\epsilon_{k\sigma})$ denote the particle and hole excitations, respectively:
\begin{align} 
\lambda_{1-\sigma}(\epsilon_{k\sigma}) =  U^{2} \tilde{\eta}^{2}
\int \frac{d\epsilon_{1}d\epsilon_{3}d\epsilon_{4} \,
\rho_{\sigma}(\epsilon_{1}) \rho_{-\sigma}(\epsilon_{3}) \rho_{-\sigma}(\epsilon_{4})
f(\epsilon_{1}) f(\epsilon_{3}) f(-\epsilon_{4})} 
{(\epsilon_{4} - \epsilon_{3} + \epsilon_{k\sigma} - \epsilon_{1} -\epsilon_{\rm c})^{2}} 
\ \ \ \  ,
\label{lambda1}
\end{align}
\begin{align} 
\lambda_{2-\sigma}(\epsilon_{k\sigma}) =  U^{2} \tilde{\eta}^{2}
\int \frac{d\epsilon_{2}d\epsilon_{3}d\epsilon_{4} \,
\rho_{\sigma}(\epsilon_{2}) \rho_{-\sigma}(\epsilon_{3}) \rho_{-\sigma}(\epsilon_{4})
f(-\epsilon_{2}) f(\epsilon_{3}) f(-\epsilon_{4})} 
{(\epsilon_{4} - \epsilon_{3} + \epsilon_{2} - \epsilon_{k\sigma} -\epsilon_{\rm c})^{2}} \  .
\label{lambda2}
\end{align}

Average quasiparticle weight $Z_{\sigma}$ is obtained from the jump of $\langle n_{k\sigma} \rangle$ at the Fermi level as follows.
\begin{eqnarray}
Z_{\sigma} = 1 -  \dfrac{\lambda_{1-\sigma}(0) + \lambda_{2-\sigma}(0)}
{1 + C U^{2}  \tilde{\eta}^{2}} \ .
\label{zk}
\end{eqnarray}
The mass enhancement factor $(m^{*}/m)_{\sigma}$ for spin $\sigma$ is obtained from the quasiparticle weight as $(m^{*}/m)_{\sigma}=1/Z_{\sigma}$.
\subsection{MLA to the quasiparticle excited states}

Let us consider the quasiparticle (QP) excited states for $N-1$ electron system in metals.  The same results of the quasiparticle states are also obtained from the $N+1$ electron system.
In the weak Coulomb interaction limit, the excitation when an electron with momentum $k$ and spin $\sigma$ is removed is given by the Hartree-Fock excited state 
$|\Phi^{(N-1)}_{k\sigma} \rangle$ as follows.
\begin{eqnarray}
|\Phi^{(N-1)}_{k\sigma} \rangle = a_{k\sigma} |\phi \rangle \ .
\label{qphf}
\end{eqnarray}
Here $|\phi \rangle$ is the Hartree-Fock ground state as has been introduced in the last subsection.
According to the Fermi liquid picture \cite{land56}, the quasiparticle excited state 
$|\Psi^{(N-1)}_{k\sigma} \rangle$ for interacting electrons with momentum $k$ and spin $\sigma$  is adiabatically connected to the Hartree-Fock excited state 
$|\Phi^{(N-1)}_{k\sigma} \rangle$ with the same momentum and spin.  This implies that the QP state is expressed by a waveoperator $\Omega$ as follows.
\begin{eqnarray}
|\Psi^{(N-1)}_{k\sigma} \rangle = \Omega \, |\Phi^{(N-1)}_{k\sigma} \rangle \ .
\label{qppsi}
\end{eqnarray}

The quasiparticle energy $E_{k\sigma}$ is defined by
\begin{eqnarray}
E_{k\sigma} = - (E_{k\sigma}(N-1) - E_{0}(N)) \ .
\label{qpener}
\end{eqnarray}
Here $E_{k\sigma}(N-1)$ is the total energy for the quasiparticle excited state given by
\begin{eqnarray}
E_{k\sigma}(N-1) = \dfrac{\langle \Psi^{(N-1)}_{k\sigma} | H |\Psi^{(N-1)}_{k\sigma} \rangle}
{\langle \Psi^{(N-1)}_{k\sigma} | \Psi^{(N-1)}_{k\sigma} \rangle} \ .
\label{qptener}
\end{eqnarray}
$E_{0}(N)$ in Eq. (\ref{qpener}) is the ground-state energy for $N$ electron system. 
\begin{eqnarray}
E_{0}(N) = \dfrac{\langle \Psi_{0} | H |\Psi_{0} \rangle}
{\langle \Psi_{0} | \Psi_{0} \rangle} \ .
\label{gener}
\end{eqnarray}
Note that the Hamiltonian in Eqs. (\ref{qptener}) and (\ref{gener}) is defined by Eq. (\ref{hub}) with the atomic level measured from the Fermi level $\epsilon_{\rm F}$.

Substituting Eqs. (\ref{qptener}) and (\ref{gener}) into Eq. (\ref{qpener}), we obtain for $\epsilon_{k\sigma} < 0$
\begin{eqnarray}
E_{k\sigma} = \epsilon_{k\sigma} 
- \dfrac{\langle \Psi^{(N-1)}_{k\sigma} | \tilde{H}_{k\sigma} |\Psi^{(N-1)}_{k\sigma} \rangle}
{\langle \Psi^{(N-1)}_{k\sigma} | \Psi^{(N-1)}_{k\sigma} \rangle} 
+ \dfrac{\langle \Psi_{0} | \tilde{H} |\Psi_{0} \rangle}
{\langle \Psi_{0} | \Psi_{0} \rangle} \ .
\label{qpener2}
\end{eqnarray}
Here $\tilde{H}_{k\sigma} = H - \langle H \rangle_{0k\sigma}$, $\tilde{H} = H - \langle H \rangle_{0}$, $\langle \sim \rangle_{0k\sigma} = \langle \Phi^{(N-1)}_{k\sigma} | (\sim) | \Phi^{(N-1)}_{k\sigma}\rangle$, and we used the relation 
$\langle \tilde{H} \rangle_{0k\sigma} = -\epsilon_{k\sigma}$.
The first term at the rhs of Eq. (\ref{qpener2}) is the Hartree-Fock independent-particle energy.  The second and third terms denote the correlation corrections to the quasiparticle energy.

Using the waveoperator (\ref{qppsi}) and that for the ground state $\Omega_{0}$,
\begin{eqnarray}
|\Psi_{0} \rangle = \Omega_{0} \, |\phi \rangle \ ,
\label{gspsi}
\end{eqnarray}
we obtain the quasiparticle energy for the $N-1$ system as follows.
\begin{eqnarray}
E_{k\sigma} = \epsilon_{k\sigma} 
- \dfrac{\langle \Omega^{\dagger} \tilde{H}_{k\sigma} \Omega \rangle_{0k\sigma}}
{\langle \Omega^{\dagger} \Omega \rangle_{0k\sigma}}
+ \dfrac{\langle \Omega^{\dagger}_{0} \tilde{H} \Omega_{0} \rangle_{0}}
{\langle \Omega^{\dagger}_{0} \Omega_{0} \rangle_{0}} \ .
\label{qpener3}
\end{eqnarray}

It should be noted that the Hartree-Fock excited states $|\Phi^{(N-1)}_{k\sigma}\rangle$ 
in Eq. (\ref{qppsi}) are obtained by removing only one electron with momentum $k$ among $N$ electrons at the ground state $|\phi \rangle$.  Therefore, the waveoperator $\Omega$ describing the correlated motion of electrons should be approximately the same as the ground-state operator $\Omega_{0}$.  Thus, we make use of the following approximation,
\begin{eqnarray}
\Omega = \Omega_{0} = \Omega_{\rm MLA} \ .
\label{appwo}
\end{eqnarray}
We call the wavefunction (\ref{qppsi}) with use of Eq. (\ref{appwo}) the MLA for the QP excited state, {\it i.e.}, the QPMLA.
Although we can choose the best variational parameters 
$\eta_{k^{\prime}_{2}k_{2}k^{\prime}_{1}k_{1}}$ for the QPMLA wavefunction with use of the variational principle for the excited states, the corrections to those at the ground state are only of the order of $1/N$, so that we can neglect them in the following calculations of the quasiparticle energy.

Making use of the approximation (\ref{appwo}), the quasiparticle energy (\ref{qpener3}) is expressed as follows.
\begin{eqnarray}
E_{k\sigma} = \epsilon_{k\sigma} 
- \dfrac{\langle \Omega^{\dagger}_{\rm MLA} \tilde{H}_{k\sigma} \Omega_{\rm MLA} \rangle_{0k\sigma}}
{\langle \Omega^{\dagger}_{\rm MLA} \Omega_{\rm MLA} \rangle_{0k\sigma}}
+ \dfrac{\langle \Omega^{\dagger}_{\rm MLA} \tilde{H} \Omega_{\rm MLA} \rangle_{0}}
{\langle \Omega^{\dagger}_{\rm MLA} \Omega_{\rm MLA} \rangle_{0}} \ .
\label{qpener4}
\end{eqnarray}
Since it is not easy to calculate the correlation terms at the rhs of Eq. (\ref{qpener4}), we  make use of the SSA (\ref{ssa}):
\begin{eqnarray}
E_{k\sigma} = \epsilon_{k\sigma} 
- L \dfrac{\langle R^{\dagger}_{i} \tilde{H}_{k\sigma} R_{i} \rangle_{0k\sigma}}
{\langle R^{\dagger}_{i} R_{i} \rangle_{0k\sigma}}
+ \langle \tilde{H} \rangle \ ,
\label{qpessa}
\end{eqnarray}
or
\begin{eqnarray}
E_{k\sigma} &=& \epsilon_{k\sigma} 
- L \Big( 1 - \dfrac{\langle R^{\dagger}_{i} R_{i} \rangle_{0}}
{\langle R^{\dagger}_{i} R_{i} \rangle_{0k\sigma}} \Big) \langle \tilde{H} \rangle_{0k\sigma} 
\nonumber \\
& & - \, \dfrac{\langle R^{\dagger}_{i} R_{i} \rangle_{0}}
{\langle R^{\dagger}_{i} R_{i} \rangle_{0k\sigma}} L \Big( 
\dfrac{\langle R^{\dagger}_{i} \tilde{H}_{0} R_{i} \rangle_{0k\sigma}}
{\langle R^{\dagger}_{i} R_{i} \rangle_{0}} - \langle \tilde{H}_{0} \rangle_{0k\sigma} 
- \frac{1}{L}  \langle \tilde{H}_{0} \rangle \Big)   \nonumber \\
& & - \dfrac{\langle R^{\dagger}_{i} R_{i} \rangle_{0}}
{\langle R^{\dagger}_{i} R_{i} \rangle_{0k\sigma}} L \Big( 
\dfrac{\langle R^{\dagger}_{i} \tilde{H}_{I} R_{i} \rangle_{0k\sigma}}
{\langle R^{\dagger}_{i} R_{i} \rangle_{0}} - \langle \tilde{H}_{I} \rangle_{0k\sigma} 
- \frac{1}{L}  \langle \tilde{H}_{I} \rangle \Big)   \nonumber \\ 
& & + L \Big( 1 - \dfrac{\langle R^{\dagger}_{i} R_{i} \rangle_{0}}
{\langle R^{\dagger}_{i} R_{i} \rangle_{0k\sigma}} \Big) \epsilon_{\rm c} \ .
\label{qpessa2}
\end{eqnarray}
Here $\langle \tilde{H} \rangle_{0k\sigma} = \langle \tilde{H_{0}} \rangle_{0k\sigma} + \langle \tilde{H_{I}} \rangle_{0k\sigma}$, 
$\langle \tilde{H} \rangle = \langle \tilde{H_{0}} \rangle + \langle \tilde{H_{I}} \rangle = L \epsilon_{\rm c}$, and $\epsilon_{\rm c}$ is the correlation energy in the SSA given by Eq. (\ref{ecssa}).
Each term at the rhs of Eq. (\ref{qpessa2}) can be calculated with use of the Wick theorem.  The results are given in Appendix.

Using the results of calculations in Appendix and the interpolated form of variational parameters (\ref{etaint}), we reach the following expression of the quasiparticle energy.
\begin{eqnarray}
E_{k\sigma} &=& \epsilon_{k\sigma} 
- \dfrac{(2-\tilde{\eta})U}{\langle R^{\dagger}_{i} R_{i} \rangle_{0}} 
\big( \chi_{2\sigma}(\epsilon_{k\sigma}) - \chi_{1\sigma}(\epsilon_{k\sigma}) \big)    \nonumber \\
& & - \dfrac{U}{\langle R^{\dagger}_{i} R_{i} \rangle_{0}} 
\int d\epsilon \rho_{-\sigma}(\epsilon) \Big[ 
f(-\epsilon) \big\{ \nu_{3\sigma}(\epsilon-\epsilon_{k\sigma})^{2} + 
\nu_{4\sigma}(\epsilon+\epsilon_{k\sigma})^{2} \big\} 
\nonumber \\
& & \hspace*{40mm} - f(\epsilon) \big\{ \nu_{2\sigma}(\epsilon+\epsilon_{k\sigma})^{2} + 
\nu_{3-\sigma}(\epsilon_{k\sigma}-\epsilon)^{2} \big\} \Big]
\nonumber \\
& & \hspace*{-5mm} - \dfrac{2U}{\langle R^{\dagger}_{i} R_{i} \rangle_{0}}
\int d\epsilon d\epsilon^{\prime} \rho_{\sigma}(\epsilon) 
\rho_{-\sigma}(\epsilon^{\prime}) f(-\epsilon)
\nonumber \\
& & \times \big[ f(\epsilon^{\prime})
\tau^{(-)}_{1\sigma}(-\epsilon^{\prime} \! + \! \epsilon \! - \! \epsilon_{k\sigma})
\nu_{3-\sigma}(\epsilon \! - \! \epsilon^{\prime})
- f(-\epsilon^{\prime})
\tau^{(+)}_{2\sigma}(\epsilon^{\prime} \! + \! \epsilon \! - \! \epsilon_{k\sigma})
\nu_{4\sigma}(\epsilon^{\prime} \! + \! \epsilon) \big]_{}
\nonumber \\ 
& & \hspace*{-5mm} + \dfrac{2U^{2}\tilde{\eta}}{\langle R^{\dagger}_{i} R_{i} \rangle_{0}}
\int \dfrac{d\epsilon_{3}d\epsilon_{4}
\rho_{-\sigma}(\epsilon_{3})\rho_{-\sigma}(\epsilon_{4})f(\epsilon_{3})f(-\epsilon_{4}) 
(\chi_{2-\sigma}(\epsilon_{3})-\chi_{1-\sigma}(\epsilon_{4}))}
{\epsilon_{4} - \epsilon_{3} - \epsilon_{\rm c}}
\nonumber \\
& & \hspace*{-5mm} - \dfrac{2U}{\langle R^{\dagger}_{i} R_{i} \rangle_{0}}
\int d\epsilon d\epsilon^{\prime} \rho_{\sigma}(\epsilon) 
\rho_{-\sigma}(\epsilon^{\prime}) f(\epsilon)
\nonumber \\
& & \hspace*{-5mm} \times \big[ f(\epsilon^{\prime})
\tau^{(-)}_{1\sigma}(-\epsilon^{\prime}+\epsilon_{k\sigma}-\epsilon)\nu_{2\sigma}(\epsilon^{\prime}+\epsilon)
- f(-\epsilon^{\prime})
\tau^{(+)}_{2\sigma}(\epsilon^{\prime}+\epsilon_{k\sigma}-\epsilon)\nu_{3\sigma}(\epsilon^{\prime}-\epsilon) \big] 
\nonumber \\
& & \hspace*{-15mm} + \dfrac{U^{2}\tilde{\eta}}{\langle R^{\dagger}_{i} R_{i} \rangle_{0}}
\int \dfrac{\displaystyle \Big[\prod^{4}_{n=1} \!\! d\epsilon_{n}\Big] 
X_{-\sigma}(\epsilon_{1}, \! \epsilon_{2}, \! \epsilon_{3}, \! \epsilon_{4}) 
\big( \tau^{(-)}_{1\sigma}(-\epsilon_{1} \! + \! \epsilon_{4} \! - \! \epsilon_{3})
- \tau^{(+)}_{2\sigma}(\epsilon_{2} \! + \! \epsilon_{4} \! - \! \epsilon_{3}) \big)}
{\epsilon_{4} - \epsilon_{3} + \epsilon_{2} - \epsilon_{1} - \epsilon_{\rm c}}  \ .
\label{qpessa3}
\end{eqnarray}
Here $\langle R^{\dagger}_{i} R_{i} \rangle_{0} = 1 + CU^{2}\tilde{\eta}^{2}$, 
$X_{\sigma}(\epsilon_{1}, \epsilon_{2}, \epsilon_{3}, \epsilon_{4})$ has been given by 
Eq. (\ref{xsigma}), and
\begin{eqnarray}
\chi_{1\sigma}(\epsilon) = U\tilde{\eta} 
\int \dfrac{d\epsilon_{1}d\epsilon_{2}d\epsilon_{3}
\rho_{-\sigma}(\epsilon_{1})\rho_{-\sigma}(\epsilon_{2})\rho_{\sigma}(\epsilon_{3}) f(\epsilon_{1})f(-\epsilon_{2})f(\epsilon_{3})}
{\epsilon - \epsilon_{3} + \epsilon_{2} - \epsilon_{1} - \epsilon_{\rm c}} \ ,
\label{chi1}
\end{eqnarray}
\begin{eqnarray}
\chi_{2\sigma}(\epsilon) = U\tilde{\eta} 
\int \dfrac{d\epsilon_{1}d\epsilon_{2}d\epsilon_{4}
\rho_{-\sigma}(\epsilon_{1})\rho_{-\sigma}(\epsilon_{2})\rho_{\sigma}(\epsilon_{4}) f(\epsilon_{1})f(-\epsilon_{2})f(-\epsilon_{4})}
{\epsilon_{4} - \epsilon + \epsilon_{2} - \epsilon_{1} - \epsilon_{\rm c}} \ ,
\label{chi2}
\end{eqnarray}
\begin{eqnarray}
\nu_{2\sigma}(\epsilon) = U\tilde{\eta} 
\int \dfrac{d\epsilon_{2}d\epsilon_{4}
\rho_{\sigma}(\epsilon_{2})\rho_{-\sigma}(\epsilon_{4}) f(-\epsilon_{2})
f(-\epsilon_{4})}
{\epsilon_{4} - \epsilon + \epsilon_{2} - \epsilon_{\rm c}} \ ,
\label{nu2}
\end{eqnarray}
\begin{eqnarray}
\nu_{3\sigma}(\epsilon) = U\tilde{\eta} 
\int \dfrac{d\epsilon_{2}d\epsilon_{3}
\rho_{\sigma}(\epsilon_{2})\rho_{-\sigma}(\epsilon_{3}) f(-\epsilon_{2})
f(\epsilon_{3})}
{\epsilon - \epsilon_{3} + \epsilon_{2} - \epsilon_{\rm c}} \ ,
\label{nu3}
\end{eqnarray}
\begin{eqnarray}
\nu_{4\sigma}(\epsilon) = U\tilde{\eta} 
\int \dfrac{d\epsilon_{1}d\epsilon_{3}
\rho_{\sigma}(\epsilon_{1})\rho_{-\sigma}(\epsilon_{3}) f(\epsilon_{1})
f(\epsilon_{3})}
{\epsilon - \epsilon_{3} - \epsilon_{1} - \epsilon_{\rm c}} \ ,
\label{nu4}
\end{eqnarray}
\begin{eqnarray}
\tau^{(-)}_{1\sigma}(\epsilon) = U\tilde{\eta} 
\int \dfrac{d\epsilon_{4} \rho_{-\sigma}(\epsilon_{4}) f(-\epsilon_{4})}
{\epsilon_{4} + \epsilon - \epsilon_{\rm c}} \ ,
\label{tau1}
\end{eqnarray}
\begin{eqnarray}
\tau^{(+)}_{2\sigma}(\epsilon) = U\tilde{\eta} 
\int \dfrac{d\epsilon_{3} \rho_{-\sigma}(\epsilon_{3}) f(\epsilon_{3})}
{\epsilon - \epsilon_{3} - \epsilon_{\rm c}} \ .
\label{tau2}
\end{eqnarray}

It should be noted that the first term at the rhs of Eq. (\ref{qpessa3}) is of the order of $U^{0}$, the second term is of the order of $U^{2}$, and the other terms are of the order of $U^{3}$ for small Coulomb interaction $U$.
Moreover, $\tilde{\eta} \rightarrow 1$ in the weak Coulomb interaction limit.
Thus, Eq. (\ref{qpessa3}) reduces to the result of the second order perturbation theory in the small Coulomb interaction limit:
\begin{eqnarray}
E_{k\sigma} = \epsilon_{k\sigma} 
- U \big( \chi_{2\sigma}(\epsilon_{k\sigma}) - \chi_{1\sigma}(\epsilon_{k\sigma}) \big)  
+ \cdots  \ . 
\label{eksecond}
\end{eqnarray}

It is convenient for the numerical calculations to make use of the Laplace transformation:
\begin{eqnarray}
\frac{1}{z-\epsilon} = -i \int^{\infty}_{0} dt \, {\rm e}^{i(z-\epsilon)t}  \ , 
\label{lapt}
\end{eqnarray}
where $z = \omega + i\delta$, $\delta$ being the infinitesimal positive-definite number.
Applying the Laplace transform (\ref{lapt}) for the denominators of the integrands at the rhs of Eq. (\ref{qpessa3}), we obtain
\begin{eqnarray}
E_{k\sigma} &=& \epsilon_{k\sigma} 
- \dfrac{(2-\tilde{\eta})U}{\langle R^{\dagger}_{i} R_{i} \rangle_{0}} 
\big( \chi_{2\sigma}(\epsilon_{k\sigma}) - \chi_{1\sigma}(\epsilon_{k\sigma}) \big)    \nonumber \\
& & \hspace*{-5mm} - \dfrac{U^{3}\tilde{\eta}^{2}}{\langle R^{\dagger}_{i} R_{i} \rangle_{0}} 
\int^{\infty}_{0} dtdt^{\prime} {\rm e}^{i\epsilon_{\rm c}(t+t^{\prime})}
\nonumber \\
& & \hspace*{-5mm} \times \big( 
{\rm e}^{-i\epsilon_{k\sigma}(t+t^{\prime})} a_{\sigma}(-t)a_{\sigma}(-t^{\prime}) +
{\rm e}^{i\epsilon_{k\sigma}(t+t^{\prime})} b_{\sigma}(t)b_{\sigma}(t^{\prime}) +
a_{\sigma}(-t-t^{\prime})b_{\sigma}(t+t^{\prime}) \big)
\nonumber \\
& & \hspace*{0mm} \times \big( 
a_{-\sigma}(-t-t^{\prime})b_{-\sigma}(t)b_{-\sigma}(t^{\prime}) -
b_{-\sigma}(t+t^{\prime})a_{-\sigma}(-t)a_{-\sigma}(-t^{\prime}) \big)
\nonumber \\
& & \hspace*{-5mm} + \dfrac{2U^{3}\tilde{\eta}^{2}}{\langle R^{\dagger}_{i} R_{i} \rangle_{0}} 
\int^{\infty}_{0} dtdt^{\prime} {\rm e}^{i\epsilon_{\rm c}(t+t^{\prime})}
\nonumber \\
& & \hspace*{-5mm} \times \big( 
{\rm e}^{-i\epsilon_{k\sigma}t} a_{\sigma}(-t-t^{\prime})b_{\sigma}(t^{\prime}) +
{\rm e}^{i\epsilon_{k\sigma}t} b_{\sigma}(t+t^{\prime})a_{\sigma}(-t^{\prime}) -
a_{\sigma}(-t^{\prime})b_{\sigma}(t^{\prime}) \big)
\nonumber \\
& & \hspace*{0mm} \times \big( 
a_{-\sigma}(-t-t^{\prime})b_{-\sigma}(t)b_{-\sigma}(t^{\prime}) -
b_{-\sigma}(t+t^{\prime})a_{-\sigma}(-t)a_{-\sigma}(-t^{\prime}) \big) \ .
\label{qpelap}
\end{eqnarray}
Here
\begin{eqnarray}
\chi_{2\sigma}(\epsilon_{k\sigma}) - \chi_{1\sigma}(\epsilon_{k\sigma})
= iU\tilde{\eta} \int^{\infty}_{0} dt {\rm e}^{i\epsilon_{\rm c}t} 
a_{-\sigma}(-t)b_{-\sigma}(t) \big( 
{\rm e}^{i\epsilon_{k\sigma}t} b_{\sigma}(t) - {\rm e}^{-i\epsilon_{k\sigma}t} a_{\sigma}(-t)
\big) \ .
\label{chilap}
\end{eqnarray}
The expressions of the other physical quantities at the ground state have been given in Appendix B of Ref. (26).

It should be noted that the quasiparticle states are well defined only in the vicinity of the Fermi level.  Thus, we linearize Eq. (\ref{qpelap}) with respect to the small energy parameter $\epsilon_{k\sigma}$, and obtain the quasiparticle energy for the QPMLA as follows.
\begin{eqnarray}
E_{k\sigma} = Z_{\sigma}\epsilon_{k\sigma} + \Delta_{\sigma} \ ,
\label{qpelin}
\end{eqnarray}
\begin{eqnarray}
Z_{\sigma} = 1 - \Lambda_{\sigma} \ ,
\label{qpwtlin}
\end{eqnarray}
\begin{eqnarray}
\Lambda_{\sigma} &=& \dfrac{1}{\langle R^{\dagger}_{i} R_{i} \rangle_{0}} \Big[ \, 
\dfrac{2-\tilde{\eta}}{\tilde{\eta}}
\big( \lambda_{1-\sigma}(0) + \lambda_{2-\sigma}(0) \big)
\nonumber \\
& & \hspace*{-8mm} - 2iU^{3}\tilde{\eta}^{2} \int^{\infty}_{0} dtdt^{\prime} 
{\rm e}^{i\epsilon_{\rm c}(t+t^{\prime})} t
\nonumber \\
& & \hspace*{-5mm}  \times \big( a_{\sigma}(-t)a_{\sigma}(-t^{\prime}) - 
b_{\sigma}(t)b_{\sigma}(t^{\prime}) - a_{\sigma}(-t-t^{\prime})b_{\sigma}(t^{\prime}) +
b_{\sigma}(t+t^{\prime})a_{\sigma}(-t^{\prime}) \big)
\nonumber \\
& & \hspace*{-5mm} \times \big( a_{-\sigma}(-t-t^{\prime})b_{-\sigma}(t)b_{-\sigma}(t^{\prime}) - 
b_{-\sigma}(t+t^{\prime})a_{-\sigma}(-t)a_{-\sigma}(-t^{\prime}) \big)
\Big] \ ,
\label{lbdlin}
\end{eqnarray}
\begin{eqnarray}
\Delta_{\sigma} &=& - \dfrac{(2-\tilde{\eta})U}{\langle R^{\dagger}_{i} R_{i} \rangle_{0}} 
\big( \chi_{2\sigma}(0) - \chi_{1\sigma}(0) \big)    
\nonumber \\
& & \hspace*{-7mm} - \dfrac{U^{3}\tilde{\eta}^{2}}{\langle R^{\dagger}_{i} R_{i} \rangle_{0}} 
\int^{\infty}_{0} dtdt^{\prime} {\rm e}^{i\epsilon_{\rm c}(t+t^{\prime})}
\nonumber \\
& & \hspace*{-2mm} \times \big( 
a_{\sigma}(-t-t^{\prime})b_{\sigma}(t+t^{\prime}) + a_{\sigma}(-t)a_{\sigma}(-t^{\prime})
+ b_{\sigma}(t)b_{\sigma}(t^{\prime}) \big)
\nonumber \\
& & \hspace*{5mm} \times \big( 
a_{-\sigma}(-t-t^{\prime})b_{-\sigma}(t)b_{-\sigma}(t^{\prime}) -
b_{-\sigma}(t+t^{\prime})a_{-\sigma}(-t)a_{-\sigma}(-t^{\prime}) \big)
\nonumber \\
& & \hspace*{-7mm} + \dfrac{2U^{3}\tilde{\eta}^{2}}{\langle R^{\dagger}_{i} R_{i} \rangle_{0}} 
\int^{\infty}_{0} dtdt^{\prime} {\rm e}^{i\epsilon_{\rm c}(t+t^{\prime})}
\nonumber \\
& & \hspace*{-2mm} \times \big( a_{\sigma}(-t-t^{\prime})b_{\sigma}(t^{\prime}) + 
b_{\sigma}(t+t^{\prime})a_{\sigma}(-t^{\prime}) -
a_{\sigma}(-t^{\prime})b_{\sigma}(t^{\prime}) \big)
\nonumber \\
& & \hspace*{5mm} \times \big( 
a_{-\sigma}(-t-t^{\prime})b_{-\sigma}(t)b_{-\sigma}(t^{\prime}) -
b_{-\sigma}(t+t^{\prime})a_{-\sigma}(-t)a_{-\sigma}(-t^{\prime}) \big) \ .
\label{dltlap}
\end{eqnarray}
Here we made use of the relations $\partial \chi_{1\sigma}(0)/\partial \epsilon_{k\sigma} 
= - \lambda_{1-\sigma}(0)/U\tilde{\eta}$ and 
$\partial \chi_{2\sigma}(0)/\partial \epsilon_{k\sigma} 
= - \lambda_{2-\sigma}(0)/U\tilde{\eta}$.
The quasiparticle energy (\ref{qpelin}) is consistent with the second order perturbation theory.  Moreover, we can verify for the half-filled band with the particle-hole symmetry that the quasiparticle weight (\ref{qpwtlin}) agrees with that was obtained from the jump of the momentum distribution function at the Fermi level, {\it i.e.}, Eq. (\ref{zk}).
\section{Numerical Results}

We have performed numerical calculations for the quasiparticle (QP) weight and energy in the paramagnetic state to examine the validity of our results. The QP energy (\ref{qpelin}) works best in infinite dimensions because we adopted the SSA.  We considered first the half-filled Hubbard model on the hypercubic lattice in infinite dimensions ($d=\infty$) \cite{muller89}.  
The density of states (DOS) for the noninteracting electron system is given by $\rho(\epsilon) = (1/\sqrt{\pi}) \exp{(-\epsilon^{2})}$.  The QP weights $Z$ calculated by various methods are summarized in Fig. 1.
\begin{figure}[htbp]
%\vspace{-1.5cm}
\begin{center}
%\hspace*{15mm}
\includegraphics [scale=1.0, angle=0]{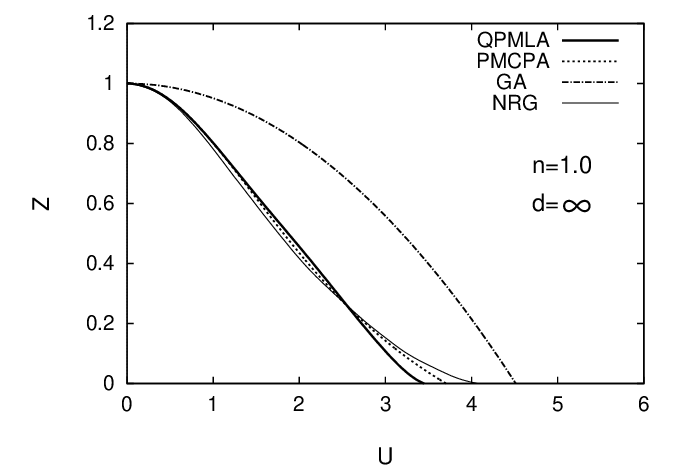}
\end{center}
%\vspace{1.5cm}
\caption{\label{figzuinfty}
Quasiparticle (QP) weight $Z$ vs Coulomb interaction energy parameter $U$ curves for the Hubbard model at half filling ($n=1.0$) on the hypercubic lattice in infinite dimensions ($d=\infty$).
Full solid curve: QP weight $Z$ in the momentum-dependent local ansatz approach to the quasiparticle excited state (QPMLA), dotted curve: QP weight in the projection operator method combined with the coherent potential approximation (PMCPA) (Ref. 40), dot-dashed curve: QP weight in the Gutzwiller ansatz wavefunction approach (GA) (Ref. 9), and thin solid curve: QP weight in the numerical renormalization group approach (NRG) (Ref. 42).
}
\end{figure}
The numerical renormalization group method (NRG) is known to describe best the infinite dimensional system \cite{bulla99}.  The QP weight in the GA \cite{br70} decreases with increasing Coulomb interaction strength $U$ as $Z({\rm GA}) = 1 - U^{2}/U_{{\rm c}2}({\rm GA})^{2}$ and vanishes at $U_{c2}({\rm GA})=8/\sqrt{\pi} (=4.51)$.  
We find that the $Z({\rm GA})$ considerably deviates upwards from the NRG curve.  The QP weight curve based on the QPMLA (, {\it i.e.}, Eq. (\ref{qpwtlin})) is close to the NRG, but the critical Coulomb interaction $U_{{\rm c}2}({\rm QPMLA})=3.40$ is somewhat smaller than the NRG value 
$U_{{\rm c}2}({\rm NRG})=4.10$.
The PMCPA is a Green function approach based on the projection operator method (PM) combined with the coherent-potential approximation (CPA) to the self-energy \cite{kake04-1}.  
The QP weight in the PMCPA is given by 
$Z({\rm PMCPA}) = (1 - U^{2}/U_{{\rm c}2}({\rm PMCPA})^{2})/(1+U^{2}/U^{2}_{2})$, where 
$U_{{\rm c}2}({\rm PMCPA})=3.71$ and $U_{2}=2.52$.  The curve yields a reasonable result which is slightly better than the QPMLA as seen in Fig. 1.  We can expect from these results that the QPMLA and the PMCPA work well in the metallic regime $U \lesssim 0.7 U_{{\rm c}2}({\rm NRG})$.

The infinite dimensional model is not available for the numerical investigations of the momentum dependence of the quasiparticle energy $E_{k\sigma}$.
We therefore adopted the Hubbard model on the simple cubic (sc) lattice with the nearest-neighbor transfer integrals $t$ for the calculations of the QP energy $E_{k\sigma}$.
\begin{figure}[htbp]
\begin{center}
%\vspace{-1.5cm}
%\hspace*{15mm}
\includegraphics [scale=1.0, angle=0]{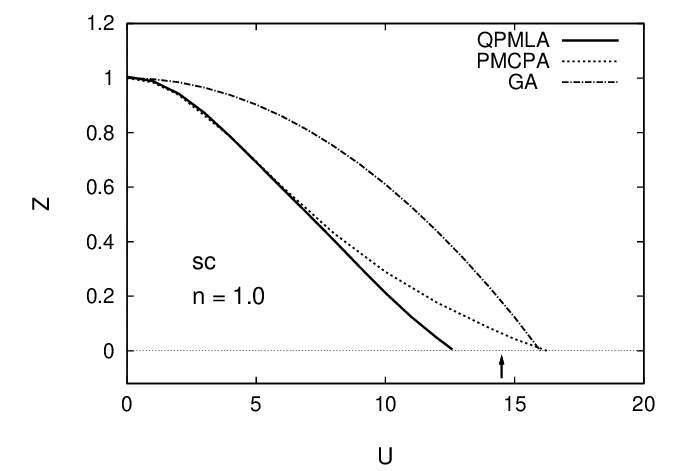}%
%\vspace{1.5cm}
\caption{\label{figzusc}
QP weight $Z$ vs $U$ curves for the Hubbard model at half filling ($n=1.0$) on the simple cubic (sc) lattice.
Full solid curve: QP weight $Z$ in the QPMLA, dotted curve: QP weight in the PMCPA, dot-dashed curve: QP weight in the GA.  The arrow indicates the critical Coulomb interaction energy parameter ($U_{{\rm c}2}$) in the dynamical mean field theory (DMFT) (Ref. 43).
}
\end{center}
\end{figure}
Figure 2 shows the QP weight $Z$ vs $U$ curves for the sc lattice in various approaches.  Basic behavior of the $Z-U$ curves are similar to those in the $d=\infty$ system (Fig. 1).
Calculated critical Coulomb interactions for the sc lattice are 
$U_{{\rm c}2}({\rm QPMLA})=12.6$,  $U_{{\rm c}2}({\rm PMCPA})=16.3$, and 
$U_{c2}({\rm GA})=16.0$ in unit of $|t|$, respectively.
The value of the dynamical mean field theory (DMFT) \cite{bulla00} corresponding to the NRG in the $d=\infty$ system is reported to be 
$U_{{\rm c}2}({\rm DMFT})=14.5$.  Comparing Fig. 2 with Fig. 1 for $d=\infty$, we expect that the QPMLA and the PMCPA yield reasonable results in the metallic region $U \lesssim 0.7 U_{{\rm c}2}({\rm DMFT}) \approx 10$ in the case of the sc lattice.
In the following, we discuss the momentum dependence of the QP energy $E_{k\sigma}$ in this metallic region.
\begin{figure}[htbp]
\begin{center}
%\vspace{-1.5cm}
%\hspace*{15mm}
\includegraphics [scale=1.0, angle=0]{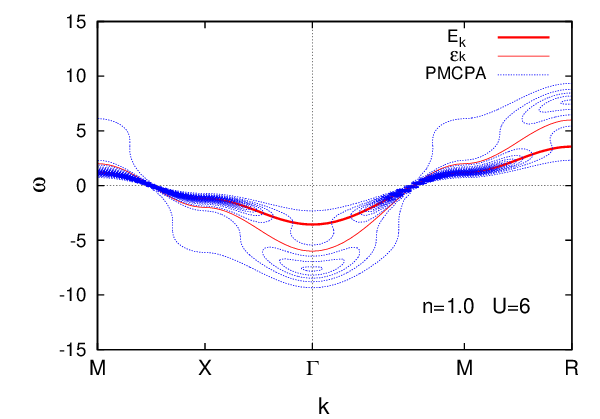}%
%\vspace{1.5cm}
\caption{\label{figqpekn1.0u6}
(Color online)
Quasiparticle (QP) energy $E_{k}$ in the QPMLA along the high-symmetry lines of the first Brillouin zone for the simple cubic lattice (solid curve (red)) and the contour map of the $k$-dependent excitation spectra obtained by the PMCPA Green function approach (dotted lines (blue)) for electron number $n=1.0$ and Coulomb interaction energy parameter $U=6$.
The Hartree-Fock (HF) energy $\epsilon_{k}$ is shown by the thin solid curve (red). 
}
\end{center}
\end{figure}

Figures $3$ shows the calculated $E_{k\sigma}$ along the high-symmetry lines of the first Brillouin zone for the sc lattice at half filling $n=1$ and Coulomb interaction energy parameter $U=6$.  The contour map of the single-particle excitation spectra $\rho(k,\omega)$ calculated by the PMCPA shows that there are well-defined lines with large amplitude crossing the Fermi level between M and X points, and between  
$\Gamma$ and M points.  These lines indicate the quasiparticle excitations.  
\begin{figure}[htbp]
\begin{center}
%\vspace{-1.5cm}
%\hspace*{15mm}
\includegraphics [scale=1.0, angle=0]{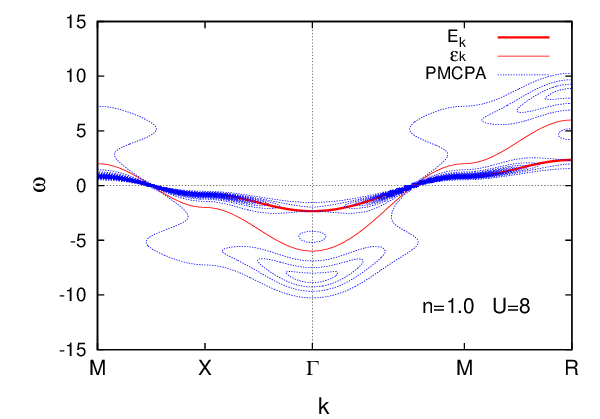}%
%\vspace{1.5cm}
\caption{\label{figqpekn1.0u10}
(Color online)
The QPMLA energy  $E_{k}$, the HF energy $\epsilon_{k}$, and the contour map of the PMCPA excitation spectra for $n=1.0$ and $U=8$.
Notation is the same as in Fig. 3.
}
\end{center}
\end{figure}
\begin{figure}[htbp]
\begin{center}
%\vspace{-1.5cm}
%\hspace*{15mm}
\includegraphics [scale=1.0, angle=0]{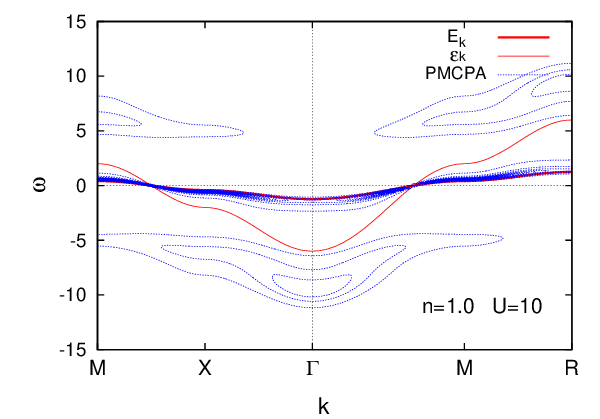}%
%\vspace{1.5cm}
\caption{\label{figqpekn1.0u8}
(Color online)
The QPMLA energy  $E_{k}$, the HF energy $\epsilon_{k}$, and the contour map of the PMCPA excitation spectra for $n=1.0$ and $U=10$.
Notation is the same as in Fig. 3.
}
\end{center}
\end{figure}

The QP energy curve obtained by the QPMLA agrees well with those obtained from the PMCPA.  The QP curve is created by strong renormalization of the Hartree-Fock energy due to electron correlations.  
Note that the QP states are well defined only in the vicinity of the Fermi level.  The QP states around $\Gamma$ point and R point disappear and incoherent excitations are dominant there according to the contour maps of the PMCPA.
Figures 4 and 5 show the QP excitation energy for larger Coulomb interactions $U=8$ and $10$.   The QP energy bands become narrow with increasing Coulomb interaction $U$ 
as expected from the $Z-U$ curve (Fig. 2).  We find again that the QPMLA curves are in good agreement with those expected from the PMCPA contour maps. 
\begin{figure}[htbp]
\begin{center}
%\vspace{-1.5cm}
%\hspace*{15mm}
\includegraphics [scale=1.0, angle=0]{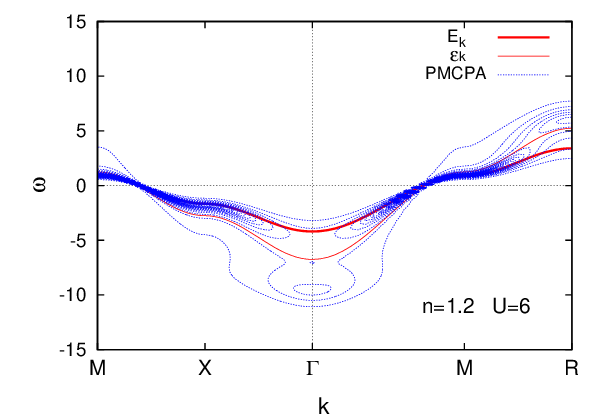}%
%\vspace{1.5cm}
\caption{\label{figqpekn1.2u6}
(Color online)
The QP energy $E_{k}$ in the QPMLA, the HF energy $\epsilon_{k}$ along the high-symmetry lines, and the contour map of the excitation spectra in the PMCPA for $n=1.2$ and $U=6$.
Notation is the same as in Fig. 3.
}
\end{center}
\end{figure}

Figures $6 \sim 8$ show the change of QP energy curves with increasing electron number per site $n$.  
The QP energy band sinks below the Fermi level with increasing electron number $n$.  
Accordingly the QP bands around $\Gamma$ and R points grow up, and the difference between the QP band and the HF one becomes smaller as the QP weight $Z$ approaches to one with increasing $n$.
We find that the QP energy curves in the QPMLA are consistent with the excitation spectra of the PMCPA also in this case.
\begin{figure}[htbp]
\begin{center}
%\vspace{-1.5cm}
%\hspace*{15mm}
\includegraphics [scale=1.0, angle=0]{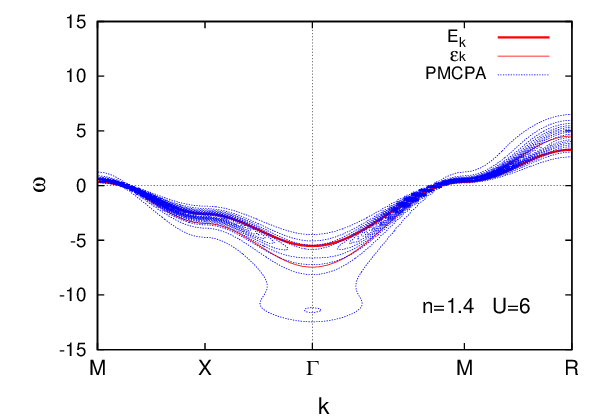}%
%\vspace{1.5cm}
\caption{\label{figqpekn1.4u6}
(Color online)
The QPMLA energy $E_{k}$, the HF energy $\epsilon_{k}$ along high-symmetry lines, and the contour map of the excitation spectra in the PMCPA for $n=1.4$ and $U=6$.
Notation is the same as in Fig. 3.
}
\end{center}
\end{figure}
\begin{figure}[htbp]
\begin{center}
%\vspace{-1.5cm}
%\hspace*{15mm}
\includegraphics [scale=1.0, angle=0]{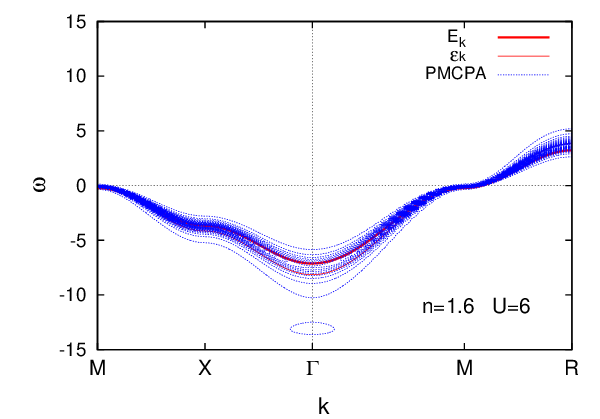}%
%\vspace{1.5cm}
\caption{\label{figqpekn1.6u6}
(Color online)
The QPMLA energy $E_{k}$, the HF energy $\epsilon_{k}$ along high-symmetry lines, and the contour map of the excitation spectra in the PMCPA for $n=1.6$ and $U=6$.
Notation is the same as in Fig. 3.
}
\end{center}
\end{figure}
\section{Summary}

We have demonstrated on the basis of the Fermi liquid picture that the quasiparticle (QP) excited states are described by the correlated wavefunction 
$|\Psi_{k\sigma} \rangle = \Omega \, a_{k\sigma}|\phi \rangle$ 
( or $\Omega \, a^{\dagger}_{k\sigma}|\phi \rangle$ ) with the ground-state waveoperator $\Omega$. 
Applying the momentum dependent local ansatz (MLA) waveoperator, 
we have derived the QP energy from the excited-state wavefunction within the SSA. 
The QP energy reduces to the second order perturbation theory in the weak Coulomb interaction limit, and is consistent with the QP weight $Z$ obtained from the jump of the momentum distribution function in the MLA.

We performed the numerical calculations for the QP energy and QP weight in the hypercubic infinite dimensional system as well as the simple-cubic system.  
We verified that the present theory describes well the QP weight of the metallic region in the high dimensional system.
The QP energy bands become narrow with increasing Coulomb interaction $U$, while they become broad when the electron number $n$ is deviated from the half filling.
We have verified that the MLA-base QP states (, {\it i.e.}, the QPMLA) describe well the QP energy bands in the high dimensional system.

One of the problems in the present approach is that the application range of the QP energy is not clarified at present, although the QP states are well defined in the vicinity of the Fermi level according to the Fermi liquid theory.  For example, we found that the QP states with large energy amplitude around $\Gamma$ and R points of the Brillouin zone for the sc lattice do not exist any more for the Coulomb interaction strength $U \gtrsim 6$ at half filling (see Fig. 3, for example).
In order to clarify the existence and stability of the QP states, one has to examine their life time $\Gamma_{k\sigma}$.  Calculations of $\Gamma_{k\sigma}$ based on the wavefunction method are left for future investigations.

The present theory is limited to the weak and intermediate Coulomb interaction energy regime.  
Extension of the QPMLA to the strongly correlated electron system is desired for understanding the low-energy excitations of correlated electrons from the viewpoint of wavefunctions.
For the quantitative description of the QP states in the strongly correlated regime, one has to take into account more the Hilbert space in the atomic region.
Improvements of the theory and extension to the realistic system are in progress.

\section*{Acknowledgment}
The author would like to express his sincere thanks to Professor Peter Fulde for his encouragements over forty years.

\appendix

\section{Calculations of Elements in Quasiparticle Energy}

In this Appendix, we present the results of calculations for each term at the rhs of the quasiparticle energy (\ref{qpessa2}). 
\begin{eqnarray}
\dfrac{\langle R^{\dagger}_{i} R_{i} \rangle_{0k\sigma}}{\langle R^{\dagger}_{i} R_{i} \rangle_{0}}
= \Big[ 
1 + \frac{1}{L} \dfrac{\lambda_{1-\sigma}(k) - \lambda_{2-\sigma}(k)}{\langle R^{\dagger}_{i} R_{i} \rangle_{0}}
\Big] f(\epsilon_{k\sigma}) \ ,
\label{riri0ks}
\end{eqnarray}
\begin{eqnarray}
\dfrac{\langle R^{\dagger}_{i} \tilde{H}_{0} R_{i} \rangle_{0k\sigma}}
{\langle R^{\dagger}_{i} R_{i} \rangle_{0}} - \langle \tilde{H}_{0} \rangle_{0k\sigma} 
- \frac{1}{L}  \langle \tilde{H}_{0} \rangle  \hspace*{-60mm}
\nonumber \\
&=& \hspace*{-2mm} \frac{1}{L} \frac{1}{\langle R^{\dagger}_{i} R_{i} \rangle_{0}} \Big[ 
2\lambda_{2-\sigma}(k)\epsilon_{k\sigma}
\nonumber \\
& & \hspace*{-7mm} + \frac{1}{L} \sum_{k^{\prime}} \big( 
\epsilon_{k^{\prime}-\sigma}\kappa_{4\sigma}(k^{\prime}, k)f(-\epsilon_{k^{\prime}-\sigma}) 
- \epsilon_{k^{\prime}-\sigma}\kappa_{3-\sigma}(k, k^{\prime})f(\epsilon_{k^{\prime}-\sigma})
- \epsilon_{k^{\prime}\sigma}\kappa_{1\sigma}(k, k^{\prime})f(\epsilon_{k^{\prime}\sigma})
\big)
\nonumber \\
& & \hspace*{-7mm} - \frac{1}{L} \sum_{k^{\prime}} \big( 
\epsilon_{k^{\prime}-\sigma}\kappa_{3\sigma}(k^{\prime}, k)f(-\epsilon_{k^{\prime}-\sigma}) 
- \epsilon_{k^{\prime}-\sigma}\kappa_{2\sigma}(k^{\prime}, k)f(\epsilon_{k^{\prime}-\sigma})
\nonumber \\
& & \hspace*{68mm} 
+ \epsilon_{k^{\prime}\sigma}\kappa_{1\sigma}(k^{\prime}, k)f(-\epsilon_{k^{\prime}\sigma})
\big) \Big] f(\epsilon_{k\sigma}) \ ,
\label{h00ks}
\end{eqnarray}
\begin{eqnarray}
\dfrac{\langle R^{\dagger}_{i} \tilde{H}_{I} R_{i} \rangle_{0k\sigma}}
{\langle R^{\dagger}_{i} R_{i} \rangle_{0}} - \langle \tilde{H}_{I} \rangle_{0k\sigma} 
- \frac{1}{L}  \langle \tilde{H}_{I} \rangle  \hspace*{-60mm}
\nonumber \\
&=& \hspace*{-2mm} \frac{1}{L} \frac{U}{\langle R^{\dagger}_{i} R_{i} \rangle_{0}} \Big[ 
2(\chi_{2\sigma}(k) -\chi_{1\sigma}(k))
\nonumber \\
& & \hspace*{-2mm}  + \frac{1}{L} \sum_{k^{\prime}} \big(
f(-\epsilon_{k^{\prime}-\sigma})\nu_{3\sigma}(k^{\prime}, k)^{2} - 
f(\epsilon_{k^{\prime}-\sigma})\nu_{2\sigma}(k^{\prime}, k)^{2}  \big)
\nonumber \\
& & \hspace*{-2mm}  + \frac{2}{L^{2}} \sum_{k^{\prime}k^{\prime\prime}}
f(-\epsilon_{k^{\prime}\sigma}) \big\{ 
f(\epsilon_{k^{\prime\prime}-\sigma})\tau^{(-)}_{1\sigma}(k^{\prime\prime}, k^{\prime}, k)\nu_{3-\sigma}(k^{\prime}, k^{\prime\prime})
\nonumber \\
& & \hspace*{50mm}  
- f(-\epsilon_{k^{\prime\prime}-\sigma})\tau^{(+)}_{2\sigma}(k^{\prime\prime}, k^{\prime}, k)\nu_{4\sigma}(k^{\prime\prime}, k^{\prime})  \big\}
\nonumber \\
& & \hspace*{-2mm}  - \frac{1}{L^{4}} \sum_{k_{1}k^{\prime}_{1}k_{2}k^{\prime}_{2}}
f(\epsilon_{k_{1}-\sigma})f(-\epsilon_{k^{\prime}_{1}-\sigma}) 
f(\epsilon_{k_{2}\sigma})f(-\epsilon_{k^{\prime}_{2}\sigma})
\eta_{k^{\prime}_{2}k_{2}k^{\prime}_{1}k_{1}}
\nonumber \\
& & \hspace*{50mm} \times \big( 
\tau^{(-)}_{1\sigma}(k_{1}, k^{\prime}_{2}, k_{2}) - 
\tau^{(+)}_{2\sigma}(k^{\prime}_{1}, k^{\prime}_{2}, k_{2})  \big) 
\nonumber \\
& & \hspace*{-2mm}  - \frac{2}{L^{2}} \sum_{k_{2}k^{\prime}_{2}}
f(\epsilon_{k_{2}-\sigma})f(-\epsilon_{k^{\prime}_{2}-\sigma})
\eta_{k^{\prime}_{2}k_{2}kk}
(\chi_{2-\sigma}(k_{2}) - \chi_{1-\sigma}(k^{\prime}_{2})) 
\nonumber \\
& & \hspace*{-2mm}  + \frac{2}{L^{3}} \sum_{k_{1}k_{2}k^{\prime}_{2}}
f(\epsilon_{k_{1}\sigma})f(\epsilon_{k_{2}-\sigma})f(-\epsilon_{k^{\prime}_{2}-\sigma})
\eta_{k^{\prime}_{2}k_{2}kk_{1}}
(\nu_{2\sigma}(k_{2}, k_{1}) - \nu_{3\sigma}(k^{\prime}_{2}, k_{1})) 
\nonumber \\
& & \hspace*{-10mm}  + \frac{1}{L^{3}} \! \sum_{k_{1}k_{2}k^{\prime}_{2}}
f(\epsilon_{k_{1}\sigma})f(\epsilon_{k_{2}-\sigma})f(-\epsilon_{k^{\prime}_{2}-\sigma})
\eta_{k^{\prime}_{2}k_{2}kk_{1}}
(\nu_{4\sigma}(k^{\prime}_{2}, k) - \nu_{3-\sigma}(k, k_{2}))  
\Big]f(\epsilon_{k\sigma}) .
\label{hi0ks}
\end{eqnarray}
Here the normalization factor $\langle R^{\dagger}_{i} R_{i} \rangle_{0} = 1 + \langle \tilde{O}^{\dagger}_{i} \tilde{O}_{i} \rangle_{0}$ has been calculated in Eq. (\ref{ecterm3}), and
\begin{eqnarray}
\lambda_{1-\sigma}(k) = 
\frac{1}{L^{3}} \sum_{k_{1}k_{2}k^{\prime}_{2}}
f(\epsilon_{k_{1}\sigma})f(\epsilon_{k_{2}-\sigma})f(-\epsilon_{k^{\prime}_{2}-\sigma})
|\eta_{k^{\prime}_{2}k_{2}kk_{1}}|^{2} \ ,
\label{lbd1app}
\end{eqnarray}
\begin{eqnarray}
\lambda_{2-\sigma}(k) = 
\frac{1}{L^{3}} \sum_{k^{\prime}_{1}k_{2}k^{\prime}_{2}}
f(-\epsilon_{k^{\prime}_{1}\sigma})f(\epsilon_{k_{2}-\sigma})f(-\epsilon_{k^{\prime}_{2}-\sigma})
|\eta_{k^{\prime}_{2}k_{2}k^{\prime}_{1}k}|^{2} \ ,
\label{lbd2app}
\end{eqnarray}
\begin{eqnarray}
\kappa_{1\sigma}(k, k^{\prime}) = 
\frac{1}{L^{2}} \sum_{k_{2}k^{\prime}_{2}}
f(\epsilon_{k_{2}-\sigma})f(-\epsilon_{k^{\prime}_{2}-\sigma})
|\eta_{k^{\prime}_{2}k_{2}kk^{\prime}}|^{2} \ ,
\label{kappa1app}
\end{eqnarray}
\begin{eqnarray}
\kappa_{2\sigma}(k, k^{\prime}) = 
\frac{1}{L^{2}} \sum_{k^{\prime}_{1}k^{\prime}_{2}}
f(-\epsilon_{k^{\prime}_{1}\sigma})f(-\epsilon_{k^{\prime}_{2}-\sigma})
|\eta_{k^{\prime}_{2}kk^{\prime}_{1}k^{\prime}}|^{2} \ ,
\label{kappa2app}
\end{eqnarray}
\begin{eqnarray}
\kappa_{3\sigma}(k, k^{\prime}) = 
\frac{1}{L^{2}} \sum_{k_{1}k^{\prime}_{2}}
f(-\epsilon_{k_{1}\sigma})f(\epsilon_{k_{2}-\sigma})
|\eta_{kk_{2}k^{\prime}_{1}k^{\prime}}|^{2} \ ,
\label{kappa3app}
\end{eqnarray}
\begin{eqnarray}
\kappa_{4\sigma}(k^{\prime}, k) = 
\frac{1}{L^{2}} \sum_{k_{1}k_{2}}
f(\epsilon_{k_{1}\sigma})f(-\epsilon_{k_{2}-\sigma})
|\eta_{k^{\prime}k_{2}kk_{1}}|^{2} \ ,
\label{kappa4app}
\end{eqnarray}
\begin{eqnarray}
\chi_{1\sigma}(k) = 
\frac{1}{L^{3}} \sum_{k_{1}k^{\prime}_{1}k_{2}}
f(\epsilon_{k_{2}\sigma})f(-\epsilon_{k^{\prime}_{1}-\sigma})f(\epsilon_{k_{1}-\sigma})
\eta_{kk_{2}k^{\prime}_{1}k_{1}} \ ,
\label{chi1app}
\end{eqnarray}
\begin{eqnarray}
\chi_{2\sigma}(k) = 
\frac{1}{L^{3}} \sum_{k_{1}k^{\prime}_{1}k^{\prime}_{2}}
f(-\epsilon_{k^{\prime}_{2}\sigma})f(-\epsilon_{k^{\prime}_{1}-\sigma})f(\epsilon_{k_{1}-\sigma})
\eta_{k^{\prime}_{2}kk^{\prime}_{1}k_{1}} \ ,
\label{chi2app}
\end{eqnarray}
\begin{eqnarray}
\nu_{2\sigma}(k^{\prime}, k) = 
\frac{1}{L^{2}} \sum_{k^{\prime}_{1}k^{\prime}_{2}}
f(-\epsilon_{k^{\prime}_{1}\sigma})f(-\epsilon_{k^{\prime}_{2}-\sigma})
\eta_{k^{\prime}_{2}k^{\prime}k^{\prime}_{1}k} \ ,
\label{nu2app}
\end{eqnarray}
\begin{eqnarray}
\nu_{3\sigma}(k^{\prime}, k) = 
\frac{1}{L^{2}} \sum_{k^{\prime}_{1}k_{2}}
f(-\epsilon_{k^{\prime}_{1}\sigma})f(\epsilon_{k_{2}-\sigma})
\eta_{k^{\prime}k_{2}k^{\prime}_{1}k} \ ,
\label{nu3app}
\end{eqnarray}
\begin{eqnarray}
\nu_{4\sigma}(k^{\prime}, k) = 
\frac{1}{L^{2}} \sum_{k_{1}k_{2}}
f(\epsilon_{k_{1}\sigma})f(\epsilon_{k_{2}-\sigma})
\eta_{k^{\prime}k_{2}kk_{1}} \ ,
\label{nu4app}
\end{eqnarray}
\begin{eqnarray}
\tau^{(-)}_{1\sigma}(k_{2}, k^{\prime}_{1}, k_{1}) = 
\frac{1}{L} \sum_{k^{\prime}_{2}}
f(-\epsilon_{k^{\prime}_{2}-\sigma})
\eta_{k^{\prime}_{2}k_{2}k^{\prime}_{1}k_{1}} \ ,
\label{tau1app}
\end{eqnarray}
\begin{eqnarray}
\tau^{(+)}_{2\sigma}(k^{\prime}_{2}, k^{\prime}_{1}, k_{1}) = 
\frac{1}{L} \sum_{k_{2}}
f(\epsilon_{k_{2}-\sigma})
\eta_{k^{\prime}_{2}k_{2}k^{\prime}_{1}k_{1}} \ .
\label{tau2app}
\end{eqnarray}

\end{document}